# Influences of a High Frequency Induction Current on the Uniformity of the Magnetic Field in an Electromagnetic Casting Mould

**Lintao Zhang, Ian Cameron and Johann Sienz**

The Advanced Sustainable Manufacturing Technologies (ASTUTE) project,
College of Engineering, Swansea University, Singleton Park, Swansea SA2 8PP,
UK
L.Zhang@swansea.ac.uk

**Abstract** An analysis of the influences of a high frequency (30 kHz) alternating current on the uniformity of the magnetic field (**B**) in an electromagnetic casting (EMC) mould is investigated by means of parametric numerical simulations where the induction current ($J_s$) varies in the range of [1 to 10000 A]. The results show that values of the magnetic flux density along the casting direction ($B_z$) near the square mould corners are small, compared to those at the other locations where $J_s$ < 10000 A, and that the magnitude of $B_z$ increases with an increased induction current ($J_s$). However, it is shown that, for the EMC mould structure investigated in this paper, the variations of $J_s$ have no significant influences on the uniformity of the magnetic field, especially for the regions near molten steel level. Moreover, the effective acting region ($R_{bz}$) for the critical magnetic field ($B_z^c$) is first introduced in this paper, which opens an interesting topic for future research.

## 1. Introduction and principles of electromagnetic casting technique

This paper focuses on investigations into the influences of the induction current ($J_s$) on the uniformities of the magnetic field (**B**) in a square electromagnetic casting (EMC) mould. The approach consists of a variety of numerical simulations to reveal the distributions of the magnetic flux density along the casting direction ($B_z$), which are obtained when the value of $J_s$ is varied.

The interest in the EMC technique, which was first applied to the aluminum casting process, stems from dramatic improvement of the surface quality of the strands[1]. The application of the EMC technique to the steel casting process has been developed in some steel casting companies[2,3]. The basic principles of EMC technique are shown in Figure 1(a). The EMC mould is surrounded by an induction coil, which is used as a carrier for the induction current (alternating current). The magnetic field (**B**) is generated in the space and the induced current ($J_i$) is also generated in the molten steel. With the interactions of **B** and $J_i$, the *Lorentz force* (**F**) acting towards the centre of the mould. The movements of the molten steel caused by the Lorentz force can improve the lubricating conditions between the strands and the mould, which result in allowing a higher casting speed together resulting in an improved production rate. Another main advantage of the EMC technique is that it can improve the surface quality of the strands so that the strands are smooth enough to be rolled without requiring the scalping process. For round billets, the surface quality is improved significantly[4], as shown in Figure 1(b). Similarly, for square billets, the depth of the oscillation marks decreases from approximately 0.6 to 0.15 mm[5].

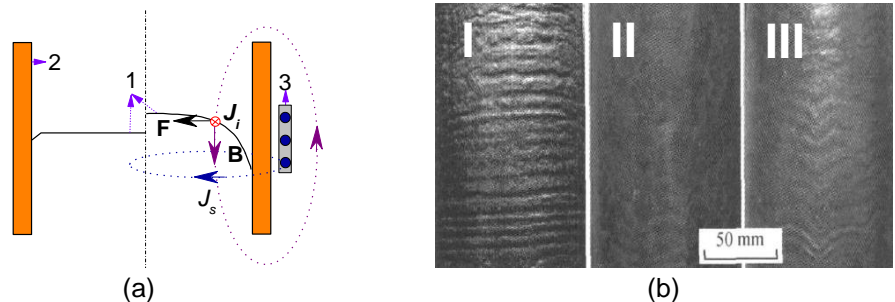

(a) (b)

Figure 1: (a): the principles of the EMC technique. Left: traditional continuous casting. Right: EMC. 1: molten steel level; 2: mold; 3: induction coil. The magnetic field **B** generated by the alternating current ($J_s$) in the induction coils. The *Lorentz force* is generated by the interactions between the induced current ($J_i$) and **B** in the liquid metal and in the mould, and acts towards the centre of the mould (b): the improvement of the surface quality of the round billets. I, II and III present the surface features when the input power is 0, 50.6 and 60 kW, respectively. The oscillation marks on the surface of 0.22wt% C steel are not present for input powers greater than approximately 50.6 kW.

Previous research into the EMC technique is widely reported in the literature, not only involving numerical simulations[6], but also industrial experiments[7]. The previous work has predominantly focused on the selection of frequency[8,9] of $J_s$, the ways to apply magnetic fields[10-12] and the influences of the mould structures on the magnetic field distributions, meniscus behavior and the surface qualities of the strands. However, from a practical point of view, the choice of the induction current (or input power) is the key problem of the industrial application of the EMC process due to economic reasons. Thus, to obtain a relatively uniform magnetic field with a wider effective range of magnetic field by using the optimum induction current for a given EMC mould structure is the dominating topic for the application of the EMC technique to commercial production. The research presented in this paper aims to determine the required input power to achieve the required magnetic field for a given mould. The approach used consists of generating the magnetic field by injecting a high frequency alternating current in the induction coil and analyzing the relationships between the current density and the uniformity of the magnetic field in the EMC mould.

The layout of this paper is as follows. The problem geometry and the numerical analysis setup are discussed in section 2. In section 3, after a short overview of $B_z$ along casting direction in the mould, the influences the molten steel simulator together with the flange on the mould top are discussed. In section 4, the influence of $J_s$ on the uniformity of $B_z$ near molten steel level are presented. The effective acting region ($R_{bz}$) for the critical magnetic field ($B_z^c$) is introduced in section 5. The main conclusions are given in section 6.

## 2. Problem geometry and numerical analysis setup

The geometry considered in this investigation is a square mould with the inner size of 100×100 mm, which consists of 12 slits, 8 large segments (LS) and 4 corner segments (CS), as shown in Figure 2.

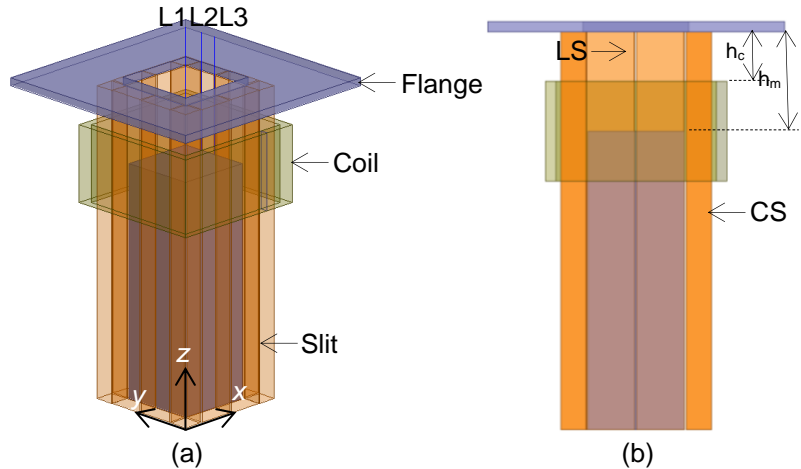

Figure 2: Three-dimensional systems for conductors and the coordinate system in the computing domain: (a) 3D view and (b) side view. The origin point is located at the bottom mould corner and the flange is placed on the top of the mould. $h_c$ is defined as the distance between the top of the induction coil and the top of the mould. $h_m$ is defined as the distance from the molten steel level to the top of the mould. In this research, $h_c$=50 mm and $h_m$=100 mm.

Unlike the geometries investigated in previous research, the sizes of the segments are not equal in this investigation. There are two main reasons for this choice: firstly, from a practical point of view, simple structures with a small number of mould segments are desirable in the applications of EMC technique in a realistic system. Secondly, as previously stated, the main objective for this paper is to investigate the influence of the induction current on the uniformity of the magnetic field in the mould, for which unequal sized segments provide increased sensitivity. In Figure 2, the origin is shown at the bottom of the mould. The flange, induction coil and coordinate system are also shown in the figure. The distance between the top of the induction coil and the top of the mould ($h_c$) is 50 mm and the distance between the top of the molten steel and the top the mould ($h_m$) is 100 mm throughout the investigations reported in this paper. The detailed parameters of the EMC mould used in the numerical simulation in this investigation are shown in Table 1.

Table1. Parameters of the EMC mould for the numerical simulation.

| Items | Values |
| --- | --- |
| Inner size of EMC mould (mm) | 100×100 |
| Height of mould (mm) | 400 |
| Height of induction coil (mm) | 100 |
| Height of liquid metal (mm) | 300 |
| Number of slits (-) | 12 |
| Height of silts (mm) | 400 |
| Width of slits (mm) | 0.5 |
| Number of corner segments (-) | 4 |
| Size of corner segment (mm) | 25×25 |
| Number of large segments (-) | 8 |
| Size of large segment (-) | 49.25×26 |
| Frequency of alternating current (kHz) | 30 |

In the whole computing domain, the Maxwell equations can be expressed as:

$$\nabla \times \mathbf{B} = \mu_0 \mathbf{J}, \quad (1)$$

$$\nabla \times \mathbf{E} = -\frac{\partial \mathbf{B}}{\partial t}, \quad (2)$$

$$\nabla \cdot \mathbf{B} = 0, \quad (3)$$

where **J**, **E**, and $\mu_0$ are current density, electric field and the vacuum permeability, respectively. Furthermore, A and ϕ are introduced as follows:

$$\mathbf{B} = \nabla \times \mathbf{A}, \quad (4)$$

$$\mathbf{E} = -\frac{\partial \mathbf{A}}{\partial t} - \nabla \phi. \quad (5)$$

This gives:

$$\nabla \times \frac{1}{\mu_0} \nabla \times \mathbf{A} - \nabla\left(\frac{1}{\mu_0} \nabla \cdot \mathbf{A}\right) + i2\pi f \sigma \mathbf{A} + \sigma \nabla \phi - \mathbf{J} = 0, \quad (7)$$

$$\nabla \cdot (-i2\pi f \sigma \mathbf{A} - \sigma \nabla \phi) = 0. \quad (8)$$

All the simulations in this paper were carried out using Ansoft Maxwell v15.0 electromagnetic field simulation software. In order to get accurate results, the meshes used were carefully generated. The skin depths for the mould (electrical conductivity $\sigma=5.96\times10^7$ S/m, permeability $\mu=1.26\times10^{-6}$ H/m) and molten steel simulator (electrical conductivity $\sigma=1.0\times10^7$ S/m, permeability $\mu=8.75\times10^{-4}$ H/m) are 0.38 mm and 2.77 mm for a 30 kHz frequency. 5 elements are controlled within the skin length. The total number of elements in the mesh for all the conductor objects is 1640763, with the mesh having increased refinement near the edges of the components. The eddy current effects of the mould and the molten steel simulator have also been taken into account. The induction current is applied to the cross-section of the induction coil as the excitation. The details of the relationship between the total energy errors and number of iterations of the numerical simulations are shown in Figure 3. In order to minimise CPU time and have reasonably accurate simulation results, all simulations were performed with 10 iterations.

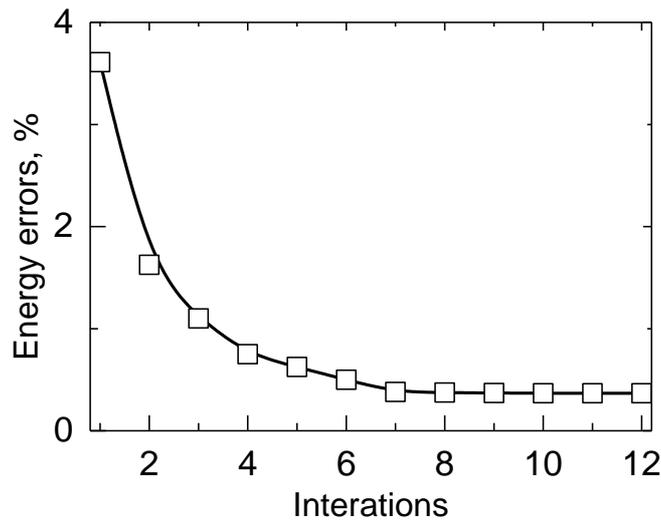

Figure 3: The relationship between the energy errors and the number of iterations of the numerical simulations, for $J_s$=1000 A.

## 3. Overview of $B_z$ along casting direction (z) in the mould

In this section, the distribution of $B_z$ along the z axis is presented for different values of $J_s$, in the range of [1 to 10000 A]. The influences of the flange, which is positioned at the top of the mould in a realistic system, and the load which is the simulation of the molten steel in the EMC mould, on the distribution of $B_z$ are also discussed. Due to the symmetry of the square EMC mould, three lines (L1, L2 and L3) on the inner surface of a single face of the inside of the mould in the casting direction (z) were selected along which to analyse the magnetic field in the mould. L1, L2 and L3 are shown in Figure 2(a), and are located near the corner of the square EMC mould, in the middle of a large segment and on the slit between two large segments, respectively.

Figure 3 shows the variations of $B_z$ along the casting direction at L1, L2 and L3 for $J_s$ = 1000 A. Three different conditions were taken into account: with load and without flange (Condition 1: green markers and lines), with load and with flange (Condition 2: blue markers and lines) and without load and without flange (Condition 3: black markers and lines). The relative positions between coil, mould, flange and load are also shown in the figure.

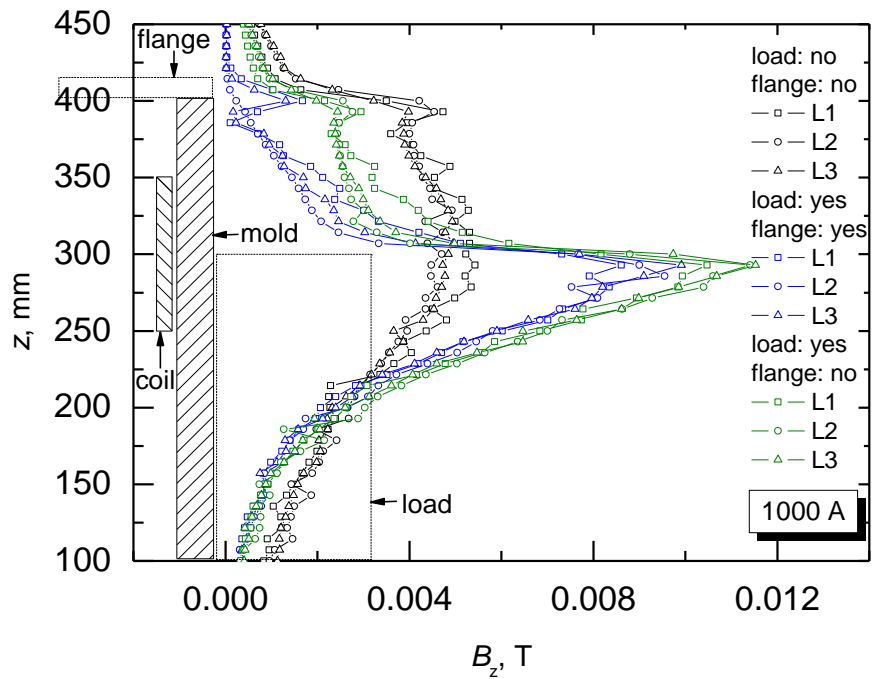

Figure 4: Comparisons of $B_z$ along z for conditions for $J_s$ = 1000 A. Condition 1: with load in the mould and without flange on top of the mould (in green). Condition 2 with load in the mould and without flange on the top of the mould (in blue). Condition 3 without load in the mould and without flange on the top of the mould. *Double peaks* of $B_z$ are shown along the casting direction for the three different conditions.

Firstly, for Condition 1, two relatively high values of $B_z$ are shown along the casting direction. The first one is located near the top of the mould (z = 400 mm) and the maximum value is located just below the molten steel level (z = 350 mm). Similar results are obtained along L1, L2 and L3. The first relatively high value is explained by the part of the magnetic field that comes into the mould through the open top of the mould. Obviously, it is more important to discuss the details of the second peak value of $B_z$, which directly influences the initial solidification process of the strands. This maximum value of $B_z$ is simply explained by the reflection of the magnetic field by the simulation of the molten steel. The influence of the molten steel causes the compression of the magnetic field, which permeates into the mould through the slits, to the inner surface of the mould. The peak value occurs just below the molten steel level. In the region between the mould top and the metal level (400 < z < 350), it is shown that values of $B_z$ near the corner region are larger than those near the middle of LS and the slit. The main reason for this is due to the two slits at the mould corner, these cause the magnetic fields to compress each other along L1. However, it is shown that the magnitude of maximum values of $B_z$ along L2 (0.0114 T at $z$ = 292 mm) and L3 (0.0115 T at $z$ = 292 mm) are similar, which are 9.5% larger than those along L1 (0.0105 T at $z$ = 292 mm). This is because the magnetic field in the region of z < 350 mm at the corner (along L1) is influenced not only by the magnetic field which permeates from the slits, but also by the induced magnetic field ($\mathbf{B_i}$, which has an opposite polarity to $\mathbf{B}$) which is generated by the eddy current in the load. The induced magnetic field at the corner region of the molten steel is higher than at the other regions of the molten steel. This causes the decrease of the total magnetic field compared to the other locations in the mould at the same height ($z$). This phenomenon indicates that for designing the EMC square mould: in order to increase $B_z$ at corner regions, it is necessary to widen the slits to some extent. In the bottom part of the mould, the $B_z$ values do not change significantly at different locations, i.e. at L1, L2 and L3.

Secondly, for Condition 2, the results show that the influences of the flange are not significant, especially for the initial solidification regions of the molten steel, as seen in Figure 4. This indicates that the magnetic field near that region is mainly formed by the magnetic field which permeates through the mould slits. However, the flange can stop the magnetic flux lines which permeate into the mould through the top, which is proven by the fact that the values of $B_z$ are lower for the Condition 2 results compared to the Condition 1 results (green in figure).

Finally, for Condition 3 the simulation also shows the influence of the molten steel. The Condition 3 results have greatly reduced peak values of Bz compared with the results from Condition 1 and Condition 2. This is simply due to the fact that the molten steel reflects the magnetic flux lines and compresses them to the inner surface of the mould, as previously discussed. This supports the fact that the effect of the molten steel must be taken into account, and will ensure accurate results for both numerical simulations and experiments.

Figure 5 shows the variations of $B_z$ along $z$, for different values of $J_s$ for Condition 1. It is shown that magnetic flux density increases as the induction current is increased, especially for the maximum values, which appear at the region just below the molten steel level. This is simply due to the fact that as the induction current increases, the *Lorentz force* increases in the mould and the air gap between the inner surface of the mould and the molten steel becomes wider. This wider gap allows an increased application of mould flux, improving the lubrication conditions. High surface quality billets will be manufactured under these conditions. However, as a result of increasing the induction current (or input power), the fluctuations in the meniscus become more pronounced, and these fluctuations can destroy the initial solidification shell, which can directly cause more defects on the surface of the strands. Obviously a greater number of defects are contrary to the initial aim of the electromagnetic continuous casting technique. This means that the optimum induction current should be selected to get the optimum surface finish for the minimum induction current or input power, and therefore minimise the power requirements. This is one of the most important topics for the application of the electromagnetic continuous casting technique.

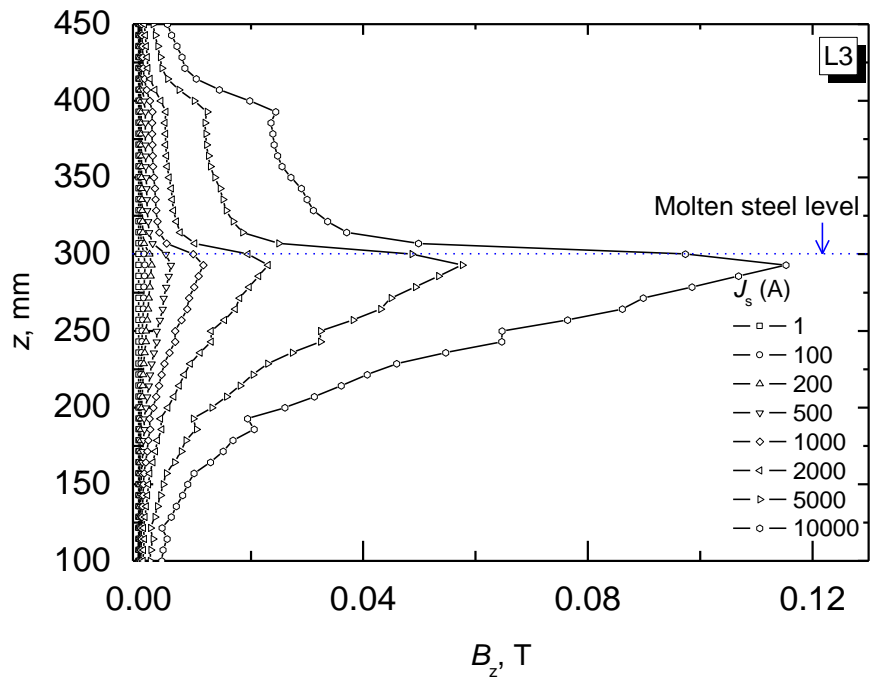

Figure 5: The variations of $B_z$ along the casting direction (L3), for different values of $J_s$. The magnitude of $B_z$ increases with increasing values of $J_s$, especially for the maximum values, which appear near the regions of the molten steel level.

## 4. Influence of $J_s$ on the uniformity of $B_z$ near molten metal level

This section first focuses on the regions where the maximum values of $B_z$ appear and the relationships between the peak value and the induction current are discussed, as shown in Figure 6. For small values of induction current, such as $J_s$ =1A, this process is effectively a quasi-traditional continuous casting (CC) process, and the amplitudes of $B_z$ are in the order of $10^{-5}$. Under this condition, the values of $B_z$ are very uniform since the effect of the induced current is negligible. As the value of $J_s$ increases, the non-uniform features become evident. For the mould investigated in this paper, it is found that, $B_z$ along L2 and L3 increases much more sharply than along L1. This phenomenon is more obvious at high values of $J_s$. However, this does not mean that the uniformity of the magnetic field becomes significantly worse as the induction current is increased. Table 2 shows the maximum values of $B_z$ along L1, L2 and L3 for different values of $J_s$. The discrepancies are defined as:

$$\varepsilon_i = \frac{B_{zmax}^{Li} - B_{zmax}^{L3}}{B_{zmax}^{L3}}, \qquad (9)$$

with i=1, 2. $B_{zmax}$ is the maximum value of $B_z$ along casting direction.

The results show that, for the geometry studied, the influence of the induction current on the uniformities of the magnetic field are not obvious for conditions where $J_s$ < 10000 A. One of the effective methods to improve the uniformity of $B_z$ is by the careful design of the corner slits, which can increase the magnetic field, as discussed in the previous sections.

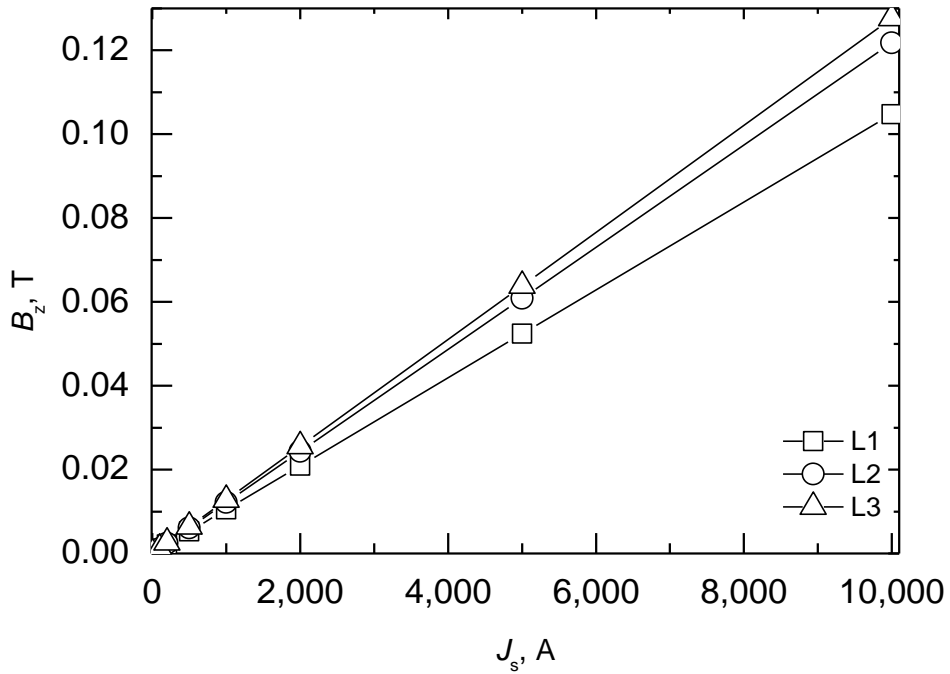

Figure 6: Comparison of $B_{zmax}$ for different values of $J_s$.

Table2. Maximum values of $B_z$ ($B_{zmax}$) along L1, L2 and L3 and their discrepancies for different values of $J_s$.

| $J_s$ (A) | $B_{zmax}$ (T) | | | $\varepsilon_1$ (%) | $\varepsilon_2$ (%) | $\varepsilon_3$ (%) |
|---|---|---|---|---|---|---|
| | L1 | L2 | L3 | | | |
| 1 | 1.05E-05 | 1.22E-05 | 1.28E-05 | 17.761 | 4.447 | - |
| 100 | 0.00105 | 0.00114 | 0.0012 | 17.896 | 4.450 | - |
| 200 | 0.00209 | 0.00244 | 0.00255 | 18.039 | 4.314 | - |
| 500 | 0.00524 | 0.00609 | 0.00638 | 17.868 | 4.545 | - |
| 1000 | 0.01047 | 0.01218 | 0.01277 | 18.010 | 4.620 | - |
| 2000 | 0.02095 | 0.02435 | 0.02554 | 17.971 | 4.659 | - |
| 5000 | 0.05237 | 0.06088 | 0.06384 | 17.966 | 4.636 | - |
| 10000 | 0.10475 | 0.12177 | 0.12768 | 17.958 | 4.628 | - |

## 5. Influence of $J_s$ on the effective acting region ($R_{Bz}$) of the critical magnetic field ($B_z^c$)

In this section, the effective acting region ($R_{Bz}$) of the critical magnetic field ($B_z^c$) is first introduced. $R_{Bz}$ is defined with the aim of capturing the effective acting range for a critical (or effective) magnetic field, which can support a sufficiently wide effective region for the EMC. This parameter is helpful for the selection of induction current and the locations of molten steel levels.

Figure 6 shows the relationships between $R_{Bz}$ and $B_z^c$, for different values of $J_s$. It is found that as $J_s$ increases, $R_{Bz}$ increases, which is also true for different values of $B_z^c$. This indicates that for higher values of induction current, the molten steel level can be designed to be at an increased range of positions compared with lower values of the induction current. It is intended to further investigate the effect on the range of satisfactory locations for the height of the molten steel, the locations of the induction coils, and their scaling laws and the results will be published in future papers.

## 6. Conclusions

A detailed analysis of magnetic flux density distributions in a square EMC mould along with the influence of the flange cover and the molten steel were investigated by varying the high frequency (30 kHz) induction alternating current. The results which were obtained from the numerical simulations have answered the question raised in the introduction:

- for the EMC mould investigated, the magnetic field near the corner region is small compared to those at the middle of the large segment and at the slit between two large segments when $J_s$ < 10000 A, at a frequency of 30 kHz.

- the increase of induction current can increase the magnitude of the magnetic field, however, it has no significant influence on the uniformity of the magnetic field in the mould, especially for the regions near the molten steel level.
- in order to obtain a relatively uniform magnetic field in the square EMC mould, the corner regions should be carefully designed for actual systems.

Furthermore, the effective acting region for the critical magnetic field was first introduced. This will form a topic of future research in this field.

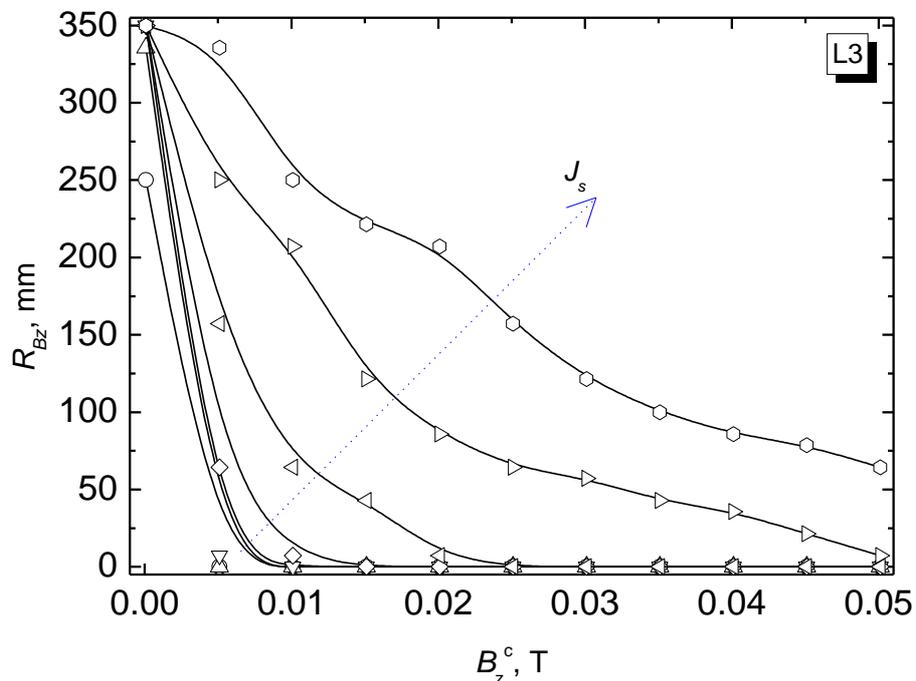

Figure 7: The variations of $R_{Bz}$ with $B_z^c$, for different values of $J_s$. The legends are the same as those in Figure 5. For a given critical magnetic field, the effective acting region increases with increasing induction current.